\pdfoutput=1
\documentclass[12pt]{article}%
\usepackage{amsmath}
\usepackage{mitpress}%
\usepackage{amsfonts}%
\usepackage{amssymb}%
\usepackage{graphicx}
\providecommand{\U}[1]{\protect\rule{.1in}{.1in}}
\newtheorem{theorem}{Theorem}

\newtheorem{definition}[theorem]{Definition}

\newtheorem{notation}[theorem]{Notation}

\newtheorem{remark}[theorem]{Remark}

\newdimen\dummy
\dummy=\oddsidemargin
\begin{document}

\title{The model of the ideal rotary element of Morita}
\author{Serban E. Vlad\\Oradea City Hall,\\P-ta Unirii 1, 410100, Oradea, ROMANIA\\email: serban\_e\_vlad@yahoo.com}
\maketitle

\section{Abstract}

Reversible computing is a concept reflecting physical reversibility. Until now
several reversible systems have been investigated. In a series of papers
Kenichi Morita defines the rotary element RE, that is a reversible logic
element. By reversibility, he understands \cite{bib10} that 'every computation
process can be traced backward uniquely from the end to the start. In other
words, they are backward deterministic systems'. He shows \cite{bib11} that
any reversible Turing machine can be realized as a circuit composed of RE's only.

Our purpose in this paper is to use the asynchronous systems theory and the
real time for the modeling of the ideal rotary element.\footnote{Mathematical
Subject Classification (2008) 94C05, 94C10, 06E30
\par
Keywords and phrases: rotary element, model, asynchronous system}

\section{Preliminaries}

\begin{definition}
The set $\mathbf{B}=\{0,1\}$ endowed with the usual algebraical laws
$\overline{\;\;},$ $\cup,\;\cdot,\;\oplus$ is called the \textbf{binary Boole
algebra}.
\end{definition}

\begin{definition}
The \textbf{characteristic function} $\chi_{A}:\mathbf{R}\rightarrow
\mathbf{B}$ of the set $A\subset\mathbf{R}$ is defined by $\forall t\in A,$%
\[
\chi_{A}(t)=\left\{
\begin{array}
[c]{c}%
1,t\in A\\
0,t\notin A
\end{array}
\right.  .
\]

\end{definition}

\begin{notation}
We denote by $Seq$ the set of the sequences $t_{k}\in\mathbf{R},$
$k\in\mathbf{N}$ which are strictly increasing $t_{0}<t_{1}<t_{2}<...$ and
unbounded from above. The elements of $Seq$ will be denoted in general by
$(t_{k}).$
\end{notation}

\begin{definition}
The \textbf{signals} are the $\mathbf{R}\rightarrow\mathbf{B}^{n}$ functions
of the form%
\begin{equation}
x(t)=\mu\cdot\chi_{(-\infty,t_{0})}(t)\oplus x(t_{0})\cdot\chi_{\lbrack
t_{0},t_{1})}(t)\oplus...\oplus x(t_{k})\cdot\chi_{\lbrack t_{k},t_{k+1}%
)}(t)\oplus...\label{pre1}%
\end{equation}
$t\in\mathbf{R},$ $\mu\in\mathbf{B}^{n},$ $(t_{k})\in Seq.$ The set of the
signals is denoted by $S^{(n)}.$
\end{definition}

\begin{definition}
In (\ref{pre1}), $\mu$ is called the \textbf{initial value} of $x$ and its
usual notation is $x(-\infty+0)$.
\end{definition}

\begin{definition}
The \textbf{left limit} of $x$ from (\ref{pre1}) is%
\[
x(t-0)=\mu\cdot\chi_{(-\infty,t_{0}]}(t)\oplus x(t_{0})\cdot\chi_{(t_{0}%
,t_{1}]}(t)\oplus...\oplus x(t_{k})\cdot\chi_{(t_{k},t_{k+1}]}(t)\oplus...
\]

\end{definition}

\begin{definition}
\label{Def10}An \textbf{asynchronous system} is a multi-valued function
$f:U\rightarrow P^{\ast}(S^{(n)}),U\in P^{\ast}(S^{(m)}).\;U$ is called the
\textbf{input set} and its elements $u\in U$ are called (\textbf{admissible})
\textbf{inputs}, while the functions $x\in f(u)$ are called (\textbf{possible}%
) \textbf{states}.
\end{definition}

\section{\label{Sec_1}The informal definition of the rotary element of Morita}

\begin{definition}
\label{Def105_}(informal) The \textbf{rotary element} RE has four inputs
$u_{1},...,$ $u_{4},$ a state $x_{0}$ and four outputs $x_{1},...,$ $x_{4}.$
Its work has been intuitively explained by the existence of a 'rotating bar'.%
%TCIMACRO{\FRAME{ftbpFU}{4.2168in}{1.9969in}{0pt}{\Qcb{RE in state
%$x_{0}(t-0)=0$ and with the input $u_{1}(t)=1$ computes $x_{0}(t)=0$ and
%$x_{1}(t)=1$}}{\Qlb{exemplu2}}{exemplu2.jpg}%
%width 4.2168in;  height 1.9969in;  depth 0pt;  original-width 4.1667in;
%original-height 1.9579in;  cropleft "0";  croptop "1";  cropright "1";
%cropbottom "0";  filename 'exemplu2.jpg';file-properties "XNPEU";}}}%
%BeginExpansion
\begin{figure}
[ptb]
\begin{center}
\includegraphics[
natheight=1.957900in,
natwidth=4.166700in,
height=1.9969in,
width=4.2168in
]%
{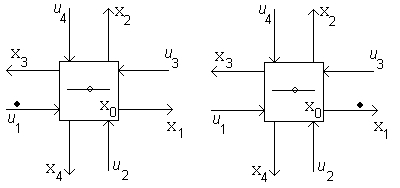}%
\caption{RE in state $x_{0}(t-0)=0$ and with the input $u_{1}(t)=1$ computes
$x_{0}(t)=0$ and $x_{1}(t)=1$}%
\label{exemplu2}%
\end{center}
\end{figure}
%EndExpansion%
%TCIMACRO{\FRAME{ftbpFU}{4.2168in}{1.9969in}{0pt}{\Qcb{RE in state
%$x_{0}(t-0)=1$ and with the input $u_{1}(t)=1$ computes $x_{0}(t)=0$ and
%$x_{4}(t)=1$}}{\Qlb{exemplu3}}{exemplu3.jpg}%
%width 4.2168in;  height 1.9969in;  depth 0pt;  original-width 4.1667in;
%original-height 1.9579in;  cropleft "0";  croptop "1";  cropright "1";
%cropbottom "0";  filename 'exemplu3.jpg';file-properties "XNPEU";}}}%
%BeginExpansion
\begin{figure}
[ptb]
\begin{center}
\includegraphics[
natheight=1.957900in,
natwidth=4.166700in,
height=1.9969in,
width=4.2168in
]%
{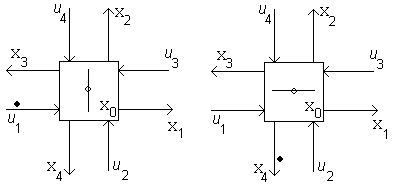}%
\caption{RE in state $x_{0}(t-0)=1$ and with the input $u_{1}(t)=1$ computes
$x_{0}(t)=0$ and $x_{4}(t)=1$}%
\label{exemplu3}%
\end{center}
\end{figure}
%EndExpansion
If (Figure \ref{exemplu2}) the state $x_{0}$ is in the horizontal position,
symbolized by us with $x_{0}(t-0)=0$, then $u_{1}(t)=1$ -this was indicated
with a bullet- makes the state remain horizontal $x_{0}(t)=0$ and the bullet
be transmitted horizontally to $x_{1},$ thus $x_{1}(t)=1.$ If (Figure
\ref{exemplu3}) $x_{0}$ is in the vertical position, symbolized by us with
$x_{0}(t-0)=1$ and if $u_{1}(t)=1$, then the state $x_{0}$ rotates
counterclockwise, i.e. it switches from $1$ to $0:x_{0}(t)=0$ and the bullet
is transmitted to $x_{4}:x_{4}(t)=1.$ No two distinct inputs may be activated
at a time -i.e. at most one bullet exists- moreover, between the successive
activation of the inputs, some time interval must exist when all the inputs
are null. If all the inputs are null, $u_{1}(t)=...=u_{4}(t)=0$ -i.e. if no
bullet exists- then $x_{0}$ keeps its previous value, $x_{0}(t)=x_{0}(t-0)$
and $x_{1}(t)=...=x_{4}(t)=0$. The definition of the rotary element is
completed by requests of symmetry.
\end{definition}

\begin{remark}
Morita states the 'reversibility' of RE. This means that in Figures
\ref{exemplu2} and \ref{exemplu3} where time passes from the left to the right
we may say looking at the right picture which the left picture is. In other
words, knowing the position of the rotating bar and the values of the outputs
allows us to know the previous position of the rotating bar and the values of
the inputs. In this 'reversed' manner of interpreting things the state $x_{0}$
rotates clockwise, $x_{1},...,x_{4}$ become inputs and $u_{1},...,u_{4}$
become outputs.

We suppose that the outputs are states, thus the state vector has the
coordinates $x=(x_{0},x_{1},x_{2},x_{3},x_{4})\in S^{(5)}.$
\end{remark}

\section{The ideal RE}

\begin{remark}
We ask that all the variables belong to $S^{(1)}$ and that any switch of the
input is transmitted to $x_{0},...,x_{4}$ instantly, without being altered and
without delays. This approximation is called by us in the following 'the ideal
RE', as opposed to 'the inertial RE'.
\end{remark}

\begin{notation}
We denote%
\[
\mathbf{0}=(0,0,0,0)\in\mathbf{B}^{4},
\]%
\[
D=\{\mathbf{0},(1,0,0,0),(0,1,0,0),(0,0,1,0),(0,0,0,1)\}.
\]

\end{notation}

\begin{definition}
The \textbf{set of the admissible inputs} $U\in P^{\ast}(S^{(4)})$ is%
\[
U=\{\lambda^{0}\cdot\chi_{\lbrack t_{0},t_{1})}\oplus\lambda^{1}\cdot
\chi_{\lbrack t_{2},t_{3})}\oplus...\oplus\lambda^{k}\cdot\chi_{\lbrack
t_{2k},\infty)}|
\]%
\[
k\in\mathbf{N},t_{0},...,t_{2k}\in\mathbf{R},t_{0}<...<t_{2k},\lambda
^{0},...,\lambda^{k}\in D\}
\]%
\[
\cup\{\lambda^{0}\cdot\chi_{\lbrack t_{0},t_{1})}\oplus\lambda^{1}\cdot
\chi_{\lbrack t_{2},t_{3})}\oplus...\oplus\lambda^{k}\cdot\chi_{\lbrack
t_{2k},t_{2k+1})}\oplus...|(t_{k})\in Seq,\lambda^{k}\in D,k\in\mathbf{N}\}.
\]

\end{definition}

\begin{notation}
$\tau^{d}:\mathbf{R}\rightarrow\mathbf{R},$ $d\in\mathbf{R}$ is the function
$\forall t\in\mathbf{R},$ $\tau^{d}(t)=t-d.$
\end{notation}

\begin{theorem}
\label{The62}The functions $u\in U$ fulfill

a) $u(-\infty+0)=\mathbf{0};$

b) $\forall i\in\{1,...,4\},$ $\forall j\in\{1,...,4\},$ $\forall
t\in\mathbf{R},$ $i\neq j$ implies%
\begin{equation}
u_{i}(t)u_{j}(t)=0,\label{fm1_}%
\end{equation}%
\begin{equation}
\overline{u_{i}(t-0)}u_{i}(t)u_{j}(t-0)\overline{u_{j}(t)}=0;\label{fm2_}%
\end{equation}

c) $\forall u\in U,$ $\forall d\in\mathbf{R,}$ $u\circ\tau^{d}\in U.$
\end{theorem}

\begin{definition}
We define the \textbf{set of the initial} (\textbf{values of the})
\textbf{states}%
\[
\Theta_{0}=\{(0,0,0,0,0),(1,0,0,0,0)\}.
\]

\end{definition}

\begin{definition}
\label{Def103}For $u\in U,$ $x\in S^{(5)},$ $x(-\infty+0)\in\Theta_{0},$ the
equations%
\begin{equation}
x_{0}(t)=\overline{x_{0}(t-0)}(u_{2}(t)\cup u_{4}(t))\cup x_{0}%
(t-0)\;\overline{u_{1}(t)}\;\overline{u_{3}(t)},\label{_re5}%
\end{equation}%
\begin{equation}
x_{1}(t)=\overline{x_{1}(t-0)}\;\overline{x_{0}(t-0)}(\overline{u_{1}%
(t-0)}u_{1}(t)\cup\overline{u_{2}(t-0)}u_{2}(t))\cup\label{_re1}%
\end{equation}%
\[
\cup x_{1}(t-0)(x_{0}(t-0)\cup\overline{u_{1}(t-0)}\cup u_{1}(t))(\overline
{x_{0}(t-0)}\cup\overline{u_{2}(t-0)}\cup u_{2}(t)),
\]%
\begin{equation}
x_{2}(t)=\overline{x_{2}(t-0)}\;x_{0}(t-0)(\overline{u_{2}(t-0)}u_{2}%
(t)\cup\overline{u_{3}(t-0)}u_{3}(t))\cup\label{_re2}%
\end{equation}%
\[
\cup x_{2}(t-0)(\overline{x_{0}(t-0)}\cup\overline{u_{2}(t-0)}\cup
u_{2}(t))(x_{0}(t-0)\cup\overline{u_{3}(t-0)}\cup u_{3}(t)),
\]%
\begin{equation}
x_{3}(t)=\overline{x_{3}(t-0)}\;\overline{x_{0}(t-0)}(\overline{u_{3}%
(t-0)}u_{3}(t)\cup\overline{u_{4}(t-0)}u_{4}(t))\cup\label{_re3}%
\end{equation}%
\[
\cup x_{3}(t-0)(x_{0}(t-0)\cup\overline{u_{3}(t-0)}\cup u_{3}(t))(\overline
{x_{0}(t-0)}\cup\overline{u_{4}(t-0)}\cup u_{4}(t)),
\]%
\begin{equation}
x_{4}(t)=\overline{x_{4}(t-0)}\;x_{0}(t-0)(\overline{u_{4}(t-0)}u_{4}%
(t)\cup\overline{u_{1}(t-0)}u_{1}(t))\cup\label{_re4}%
\end{equation}%
\[
\cup x_{4}(t-0)(\overline{x_{0}(t-0)}\cup\overline{u_{4}(t-0)}\cup
u_{4}(t))(x_{0}(t-0)\cup\overline{u_{1}(t-0)}\cup u_{1}(t))
\]
are called the \textbf{equations of the ideal RE} (of Morita) and the system
$f:U\rightarrow P^{\ast}(S^{(5)})$ that is defined by them is called the
\textbf{ideal RE}.
\end{definition}

\begin{remark}
The system $f$ is finite, i.e. $\forall u\in U,f(u)$ has two elements
$\{x,x^{\prime}\}$ satisfying $x(-\infty+0)=(0,0,0,0,0)$ and $x^{\prime
}(-\infty+0)=(1,0,0,0,0).$
\end{remark}

\begin{notation}
Let be $\mu\in\Theta_{0}.$ We denote by $f_{\mu}:U\rightarrow S^{(5)}$ the
uni-valued (i.e. deterministic) system $\forall u\in U,$%
\[
f_{\mu}(u)=x
\]
where $x$ fulfills $x(-\infty+0)=\mu$ and (\ref{_re5}),...,(\ref{_re4}).
\end{notation}

\section{\label{Sec_4}The analysis of the ideal RE}

\begin{definition}
We define $\Phi:\mathbf{B}^{5}\times\mathbf{B}^{4}\rightarrow\mathbf{B}^{5}$
by $\forall(\mu,\lambda)\in\mathbf{B}^{5}\times\mathbf{B}^{4},$ $\Phi_{0}%
(\mu,\lambda)=(\overline{\mu_{0}}(\lambda_{2}\cup\lambda_{4})\cup\mu
_{0}\overline{\lambda_{1}}\;\overline{\lambda_{3}},\overline{\mu_{0}}%
(\lambda_{1}\cup\lambda_{2}),\mu_{0}(\lambda_{2}\cup\lambda_{3}),\overline
{\mu_{0}}(\lambda_{3}\cup\lambda_{4}),\mu_{0}(\lambda_{4}\cup\lambda_{1})).$
\end{definition}

\begin{notation}
For all $k\in\mathbf{N,}$ $\lambda^{0},...,\lambda^{k},\lambda^{k+1}\in D $
and for any $\mu\in\Theta_{0},$ the vectors $\Phi(\mu,\lambda^{0}%
...\lambda^{k}\lambda^{k+1})\in\mathbf{B}^{5}$ are iteratively defined by%
\[
\Phi(\mu,\lambda^{0}...\lambda^{k}\lambda^{k+1})=\Phi(\Phi(\mu,\lambda
^{0}...\lambda^{k}),\lambda^{k+1}).
\]

\end{notation}

\begin{remark}
The iterates $\Phi(\mu,\lambda^{0}...\lambda^{k})$ show how $\Phi$ acts when a
succession of input values $\lambda^{0},...,\lambda^{k}\in D$ is applied in
the initial state $\mu\in\Theta_{0}.$ For example we have%
\[
\Phi(\mu,\mathbf{0})=\mu,
\]%
\[
\Phi(\mu,\lambda\mathbf{0}\lambda^{\prime})=\Phi(\mu,\lambda\lambda^{\prime})
\]
for any $\mu\in\Theta_{0}$ and $\lambda,\lambda^{\prime}\in D.$
\end{remark}

\begin{theorem}
\label{The64}When $\mu\in\Theta_{0},$ $\lambda,\lambda^{0},...,\lambda
^{k},...\in D$ and $(t_{k})\in Seq,$ the following statements are true:%
\begin{equation}
f_{\mu}(\lambda^{0}\cdot\chi_{\lbrack t_{0},t_{1})}\oplus\lambda^{1}\cdot
\chi_{\lbrack t_{2},t_{3})}\oplus...\oplus\lambda^{k}\cdot\chi_{\lbrack
t_{2k},\infty)})=\label{644.9}%
\end{equation}%
\[
=\mu\cdot\chi_{(-\infty,t_{0})}\oplus\Phi(\mu,\lambda^{0})\cdot\chi_{\lbrack
t_{0},t_{1})}\oplus\Phi(\mu,\lambda^{0}\mathbf{0})\cdot\chi_{\lbrack
t_{1},t_{2})}\oplus\Phi(\mu,\lambda^{0}\lambda^{1})\cdot\chi_{\lbrack
t_{2},t_{3})}\oplus...
\]%
\[
...\oplus\Phi(\mu,\lambda^{0}...\lambda^{k-1}\mathbf{0})\cdot\chi_{\lbrack
t_{2k-1},t_{2k})}\oplus\Phi(\mu,\lambda^{0}...\lambda^{k})\cdot\chi_{\lbrack
t_{2k},\infty)},
\]%
\begin{equation}
f_{\mu}(\lambda^{0}\cdot\chi_{\lbrack t_{0},t_{1})}\oplus\lambda^{1}\cdot
\chi_{\lbrack t_{2},t_{3})}\oplus...\oplus\lambda^{k}\cdot\chi_{\lbrack
t_{2k},t_{2k+1})}\oplus...)=\label{645.0}%
\end{equation}%
\[
=\mu\cdot\chi_{(-\infty,t_{0})}\oplus\Phi(\mu,\lambda^{0})\cdot\chi_{\lbrack
t_{0},t_{1})}\oplus\Phi(\mu,\lambda^{0}\mathbf{0})\cdot\chi_{\lbrack
t_{1},t_{2})}\oplus\Phi(\mu,\lambda^{0}\lambda^{1})\cdot\chi_{\lbrack
t_{2},t_{3})}\oplus...
\]%
\[
...\oplus\Phi(\mu,\lambda^{0}...\lambda^{k-1}\mathbf{0})\cdot\chi_{\lbrack
t_{2k-1},t_{2k})}\oplus\Phi(\mu,\lambda^{0}...\lambda^{k})\cdot\chi_{\lbrack
t_{2k},t_{2k+1})}\oplus...
\]

\end{theorem}

\begin{theorem}
\label{The65}$\forall\mu\in\Theta_{0},\forall u\in U,f_{\mu}(u)\in
S^{(1)}\times U.$
\end{theorem}

\begin{theorem}
\label{The71}a) $\forall\mu\in\Theta_{0},\forall\mu^{\prime}\in\Theta
_{0},\forall u\in U,$%
\[
\mu\neq\mu^{\prime}\Longrightarrow f_{\mu}(u)\neq f_{\mu^{\prime}}(u);
\]

b) $\forall\mu\in\Theta_{0},\forall u\in U,\forall u^{\prime}\in U,$%
\[
u\neq u^{\prime}\Longrightarrow f_{\mu}(u)\neq f_{\mu}(u^{\prime});
\]

c) $\forall u\in U,\forall u^{\prime}\in U,$%
\[
u\neq u^{\prime}\Longrightarrow f(u)\cap f(u^{\prime})=\emptyset.
\]

\end{theorem}

\begin{remark}
The previous Theorem states some injectivity properties of $f$. The
surjectivity property%
\[
\forall x\in S\times U,\exists\mu\in\Theta_{0},\exists u\in U,f_{\mu}(u)=x
\]
is not true$.$

Similarly with $f$, we can define $f^{-1}:U\rightarrow P^{\ast}(S^{(5)}),$
that has analugue properties with $f$. For example $\Phi^{-1}:\mathbf{B}%
^{5}\times\mathbf{B}^{4}\rightarrow\mathbf{B}^{5}$ is defined by $\forall
(\nu,\delta)\in\mathbf{B}^{5}\times\mathbf{B}^{4},\Phi_{0}^{-1}(\nu
,\delta)=(\overline{\nu_{0}}(\delta_{2}\cup\delta_{4})\cup\nu_{0}%
\overline{\delta_{1}}\;\overline{\delta_{3}},\overline{\nu_{0}}(\delta_{4}%
\cup\delta_{1}),\nu_{0}(\delta_{1}\cup\delta_{2}),\overline{\nu_{0}}%
(\delta_{2}\cup\delta_{3}),\nu_{0}(\delta_{3}\cup\delta_{4})).$

The system $f^{-1}\circ f:U\rightarrow P^{\ast}(S^{(6)})$ defined by $\forall
u\in U,$%
\[
(f^{-1}\circ f)(u)=\{(x_{0},v_{0},v_{1},v_{2},v_{3},v_{4})|x\in f(u),v\in
f^{-1}(x_{1},x_{2},x_{3},x_{4})\}
\]
does not fulfill the property $\forall u\in U,$ $\forall(x_{0},v_{0}%
,v_{1},v_{2},v_{3},v_{4})\in(f^{-1}\circ f)(u),$%
\[
u_{1}=v_{1},u_{2}=v_{2},u_{3}=v_{3},u_{4}=v_{4}%
\]
thus the conclusion of the present study is expressed by the fact that the
only 'reversibility' character of $f$ is given by its injectivity. On the
other hand, the model given by (\ref{_re5}),...,(\ref{_re4}) is reasonable,
since it satisfies non-anticipation and time invariance \cite{bib7} properties.
\end{remark}


\begin{thebibliography}{9}                                                                                                %
\bibitem {bib11}K. Morita, A simple universal logic element and cellular
automata for reversible computing, Lecture Notes in Computer Science, Springer
Berlin/Heidelberg, \textit{Machines, Computations and Universality}, Vol.
2055, 102-113, (2001).

\bibitem {bib10}K. Morita, Reversible computing and cellular automata - a
survey, \textit{Theoretical Computer Science}, Vol 395, Issue 1, 101-131, (2008).

\bibitem {bib7}S. E. Vlad, ''Teoria sistemelor asincrone'', ed. Pamantul,
Pitesti, (2007).
\end{thebibliography}
\end{document}